1   **Failure patterns caused by localized rise in pore-fluid overpressure and effective strength**

2   **of rocks**

4   A.Y. Rozhko[1], Y.Y. Podladchikov[1], and F. Renard[1,2]

5   [1]Physics of Geological Processes, University of Oslo, PO box 1048 Blindern, 0316

6   Oslo, Norway

7   [2]LGCA-CNRS-OSUG, University of Grenoble, BP 53, 38041 Grenoble, France

8   **Abstract.** In order to better understand the interaction between pore-fluid overpressure

9   and failure patterns in rocks we consider a porous elasto-plastic medium in which a

10  laterally localized overpressure line source is imposed at depth below the free surface.

11  We solve numerically the fluid filtration equation coupled to the gravitational force

12  balance and poro-elasto-plastic rheology equations. Systematic numerical simulations,

13  varying initial stress, intrinsic material properties and geometry, show the existence of

14  five distinct failure patterns caused by either shear banding or tensile fracturing. The

15  value of the critical pore-fluid overpressure $p^c$ at the onset of failure is derived from

16  an analytical solution that is in excellent agreement with numerical simulations. Finally,

17  we construct a phase-diagram that predicts the domains of the different failure patterns

18  and $p^c$ at the onset of failure.

## 1. Introduction

The effect of the *homogeneous* pore-pressure increase on the strength of crustal rocks and failure modes has been studied by many authors e.g. [*Terzaghi,* 1923; *Skempton,* 1961; *Paterson and Wong,* 2005]. Their results show that, provided that the rocks contain a connected system of pores, failure is controlled by the Terzaghi's effective stress defined as

$$\sigma'_{ij} = \sigma_{ij} - p\delta_{ij} \qquad (1)$$

where $\sigma_{ij}$ is the total stress; $p$ is the pore fluid pressure, and $\delta_{ij}$ is the Kronecker delta (by convention, compressive stress is positive).

However, many geological systems, such as magmatic dykes, mud volcanoes, hydrothermal vents, or fluid in faults, show evidence that pore pressure increase might be *localized*, instead of being *homogeneously* distributed [*Jamtveit et al.,* 2004]. Localized pore-pressure variations couple pore-fluid diffusion to rock deformation through the seepage force generated by pressure gradients [*Rice and Cleary,* 1976]. The seepage force introduces localized perturbation of the effective stress field and may promote various failure patterns. The effect of seepage forces caused by laterally homogeneous pore-pressure increase on failure patterns was recently studied experimentally by [*Mourgues and Cobbold,* 2003]. In the present study, we explore both numerically and analytically how an essentially two-dimensional, i.e. localized both at depth and laterally, increase in pore-pressure affects failure patterns in porous elasto-plastic rocks. In section 2, we discuss the effect of localized pore pressure increase on tensile and shear failure. Section 3 is devoted to the characterization of the various failure patterns using finite element and finite difference simulations that solve

45  the gravitational force balance equation and the fluid filtration equation in a poro-
46  elasto-plastic medium. In section 4, we predict the fluid pressure at the onset of failure
47  using new analytical solutions. Finally we discuss the geological implications in section
48  5.

49  **2. Effect of pore pressure on rock failure**

50  In nature, rock failure occurs in two different modes: shear bands and tensile fractures.
51  Laboratory triaxial experiments show that the Mohr-Coulomb criterion (eq. 2) provides
52  an accurate prediction for shear failure [*Paterson and Wong*, 2005]:

53  $$\tau - \sigma'_m \sin(\phi) = C \cos(\phi) \qquad (2)$$

54  where $\tau = \sqrt{(\sigma_{xx} - \sigma_{yy})^2/4 + \sigma_{xy}^2}$ is the stress deviator, $\sigma'_m = \dfrac{\sigma_{xx} + \sigma_{yy}}{2} - p$ is the mean
55  effective stress, $C$ is the rock cohesion and $\phi$ is the internal friction angle.
56  On the other hand, Griffith's theory provides a theoretical criterion for tensile failure of
57  a fluid-filled crack [*Murrell*, 1964]:

58  $$\tau - \sigma'_m = \sigma_T \qquad (3)$$

59  where $\sigma_T$ is the tensile strength of the rock. This criterion has also been verified
60  experimentally [*Jaeger*, 1963].
61  In Figure 1, we show how a homogeneous or localized increase of pore-fluid pressure
62  influences rock failure. There, $lm$ is the Mohr-Coulomb envelope (eq. 2) and $kl$ is the
63  tensile cut-off limit (eq. 3). The Mohr circle indicates the initial state of stress, with
64  zero pore-fluid overpressure. As the pore-fluid pressure increases homogeneously by an
65  amount $p$, the radius of the Mohr circle remains constant and the circle is displaced to

66 the left until it touches the failure envelope (blue curve). Depending on the location
67 where the circle touches the failure envelope, the formation of shear bands or tensile
68 fractures takes place. In both cases, when pore-fluid pressure increase is homogeneous,
69 the orientation and onset of failure patterns can be predicted [*Paterson and Wong,*
70 *2005*]. Shear bands form at an angle of $\frac{\pi}{4} - \frac{\phi}{2}$ to the direction of maximum
71 compressive stress; tensile fractures develop perpendicularly to the direction of
72 maximum tensile stress.
73 We explore a more complex scenario, where the pore-fluid increase is localized into a
74 narrow source, so that seepage forces modify locally the stress-state. As shown on
75 Figure 1c-d, the radius of the initial Mohr circle does not remain constant, as for the
76 homogeneous pore-fluid pressure increase case. For a localized fluid pressure increase
77 equal to $p$, the radius of the Mohr circle is changed by an amount of $\beta_\tau p$, and the
78 center of the circle is displaced to the left by an amount of $\beta_\sigma p$. The two
79 dimensionless parameters $\beta_\tau$ and $\beta_\sigma$ are derived below both by numerical and
80 analytical means and given by (eqs. 14-15).

81 **3. Numerical model and numerical results**

82 We consider a 2D porous medium embedded in a box of length $L$ and height $h \ll L$
83 (Figure 2a). At the bottom of this box, we define a pore-fluid over-pressure source of
84 width $w \ll h$. We consider plane-strain deformation in a material with constant and
85 homogeneous intrinsic properties. We solve numerically the fluid filtration equation
86 and the force balance equation, using a poro-elasto-plastic rheology relationship

87  between the stress and the strain rates. At initial conditions the deformation of the
88  material and the pore-fluid overpressure are equal to zero everywhere in the system.
89  The initial mechanical state is chosen below the failure limits everywhere in the system.
90  The initial vertical stress $\sigma_V$ (along y axis) is equal to the weight of the overburden.
91  The initial horizontal stress $\sigma_H$ is proportional to the vertical stress:

92
$$\begin{aligned}\sigma_V &= -\rho g y \\ \sigma_H &= A \sigma_V\end{aligned}, \quad (4)$$

93  where $\rho$ is the total density of the rock including pore fluid, $g$ is the gravitational acceleration,
94  $-y$ is the depth and $A$ is a constant coefficient. The initial shear stress is zero everywhere.
95  The top boundary ($y=0$) has a free surface condition, with zero overpressure ($p^f = 0$). The
96  lateral and bottom walls are fixed, with free-slip condition (including the pore-fluid source),
97  and impermeable (excluding pore-fluid source). The boundary conditions on lateral walls
98  represent far-field fluid pressure and mechanical state undisturbed by the localized fluid
99  pressure at the center of the model, which is achieved by using large enough horizontal extent
100 of the models, $L \gg \max(h, w)$. Perturbations of stress and displacement are negligibly small at
101 the lateral boundaries, thus either fixed stress or fixed displacement lateral boundary conditions
102 lead to the similar numerical results. At time $t = 0$, the fluid pressure is slowly increased
103 everywhere on the source segment ($p^f = p(t)$), until failure nucleates and propagates. The
104 process of fluid overpressure build up at a small segment of lower boundary is unspecified but
105 assumed slow compared to the characteristic time for establishing a steady-state distribution of
106 the fluid pressure. This quasi-static slowly driven evolution of the pressure field in a domain
107 with constant permeability is governed by the Laplace equation

108 $$\nabla^2 p^f = 0.$$ (5)

109 The gravitational force balance equation formulated for the total stress is given by

110 $$\begin{cases} \dfrac{\partial \sigma_{xx}}{\partial x} + \dfrac{\partial \sigma_{xy}}{\partial y} = 0 \\ \dfrac{\partial \sigma_{yy}}{\partial y} + \dfrac{\partial \sigma_{yx}}{\partial x} + \rho g = 0 \end{cases}$$ (6)

111 The initial state of the stress defined in (eq. 4) fulfills this relationship.

112 Using the general approach for poro-elasto-plastic deformation [*Rice and Cleary*, 1976;

113 *Vermeer*, 1990], the full strain rate tensor is given by

114 $$\dot{\varepsilon}_{ij} = \dot{\varepsilon}_{ij}^{pe} + \dot{\varepsilon}_{ij}^{pl}$$ (7)

115 where the superscripts *pe* and *pl* denote the poro-elastic and the plastic components,

116 respectively. The poro-elastic constitutive relation can be written as:

117 $$\sigma_{ij} = 2G\varepsilon_{ij}^{pe} + 2G\varepsilon_{kk}^{pe}\frac{v}{1-2v}\delta_{ij} + \alpha p^f \delta_{ij}$$ (8)

118 where $\alpha$ is the Biot-Willis poro-elastic coupling constant [*Paterson and Wong*, 2005],

119 $v$ is the drained Poisson's ratio, and $G$ is the shear modulus. The plastic strain rates

120 are given by

121 $$\begin{aligned} \dot{\varepsilon}_{ij}^{pl} &= 0 \text{ for } f < 0 \text{ or } (f = 0 \text{ and } \dot{f} < 0) \\ \dot{\varepsilon}_{ij}^{pl} &= \lambda \frac{\partial q}{\partial \sigma'_{ij}} \text{ for } f = 0 \text{ and } \dot{f} = 0 \end{aligned}$$ (9)

122 Here, we chose the yield function in the form $f = \max(f_{tension}, f_{shear})$, where $f_{tension}$ and

123 $f_{shear}$ are yield functions for failure in tension and in shear, respectively, defined as:

124 $$\begin{aligned} f_{tension} &= \tau - \sigma'_m - \sigma_T \\ f_{shear} &= \tau - \sigma'_m \sin(\phi) - C\cos(\phi) \end{aligned}$$ (10)

125 The parameter $\lambda$ in (eq. 9) is the non-negative multiplier of the plastic loading
126 [*Vermeer*, 1990], and $q$ is the plastic flow function, defined as follows for tensile
127 (associated flow rule) and shear failure (non-associated flow rule), respectively:

$$q_{tension} = \tau - \sigma'_m$$
$$q_{shear} = \tau - \sigma'_m \sin(\upsilon) \qquad (11)$$

129 where $\upsilon$ is the dilation angle ($\upsilon < \phi$). Note that the total stress is used in (eqs. 6,8),
130 whereas the Terzaghi's effective stress (eq. 1) applies in the failure equations (9-11).
131 Substitution of stresses (eq. 8) into the force balance equation (6) renders gradient of
132 the fluid pressure, commonly referred as *seepage forces*, as a cause of the solid
133 deformation.
134 Solving this set of equations, we aim to predict the localization and quasi-static
135 propagation of plastic deformations into either shear bands or tensile fractures. The
136 term tensile fracture is used here to describe the inelastic material response in the
137 process zone area that accompanies fracture onset and propagation [*Ingraffea,* 1987].
138 In order to check the independence of the simulation results on the numerical method,
139 we have developed two codes (finite element and finite difference). Extensive
140 numerical comparisons indicate that the results converge to the same values when
141 increasing the grid resolution. The poro-elastic response of codes was tested using a
142 new analytical solution (see Auxiliary Materials). The plastic response of the code was
143 tested by stretching or squeezing of lateral walls. The numerical results were consistent
144 with both numerical [*Poliakov et al.,* 1993] and laboratory experiments of rock
145 deformation [*Paterson and Wong*, 2005].

146   In the case when the lateral walls are fixed, but the localized pore-fluid overpressure
147   increases, the simulations show that the rock starts swelling poro-elastically (Figure
148   2b). When the pore-fluid pressure exceeds a critical threshold value $p^c$ at the injection
149   source, the homogeneous deformation evolves into a pattern where either a tensile
150   fracture or highly localized shear bands nucleate and propagate in a quasi-static manner
151   (Figure 2c).
152   Solving equations (5-11), the model selects the failure mode (shear or tensile) and the
153   propagation direction. Systematic numerical simulations show the existence of five
154   distinct failure patterns when the pore-fluid pressure exceeds $p^c$ (Figure 3, see
155   Auxiliary Materials for animations). Deformation patterns I (normal faulting) and II
156   (reverse faulting) form by shear failure in compressive ($\sigma_V < \sigma_H$) and in extensional
157   initial stress states ($\sigma_V > \sigma_H$), respectively. Patterns III (vertical fracturing) and IV
158   (horizontal fracturing) are caused by tensile failure in compressive and extensional
159   initial stress states, respectively. The nucleation of failure for patterns I-IV is located at
160   the fluid overpressure source. It is located at the free surface for pattern V, which is
161   also called *soil-piping* mode in hydrology [*Jones*, 1971]. After nucleation at the free
162   surface as tensile fracture in response to swelling caused by the fluid pressure build up,
163   pattern V develops by downwards propagation of a tensile failure.

164   **4. Analytical solution and failure pattern phase diagram**

165   We have derived an analytical solution for pre-failure stress distribution caused by
166   seepage forces sharing the same set of governing parameters as our numerical setup but
167   a different geometry of the outer free surface boundary [see Auxiliary Materials]. We

168 report an excellent agreement between numerical and analytical predictions of the

169 maximum pre-failure pore-fluid pressure. In order to calculate this critical pore-fluid

170 pressure $p^c$, we consider the stress and failure conditions at the fluid source (patterns I-

171 IV in Figure 3) and at the free surface (pattern V in Figure 3).

172 It follows from the analytical solution that the center, $\sigma_m$, and the radius, $\tau$, of the

173 Mohr circle do not vary along the fluid source segment. They are related by the

174 following expressions to the initial (and far field or "global') stresses and to the fluid

175 overpressure at the localized source [Auxiliary Materials]:

176 $$\tau = \left| \frac{\sigma_V - \sigma_H}{2} - \beta_\tau p \right| \qquad (12)$$

177 $$\sigma'_m = \sigma_m - p = \frac{\sigma_V + \sigma_H}{2} - \beta_\sigma p \qquad (13)$$

178 where two parameters $\beta_\sigma$ and $\beta_\tau$ control the shift of Mohr circle and radius change,

179 respectively (Figure 1):

180 $$\beta_\sigma = 1 - \frac{\alpha}{2} \frac{1-2\nu}{1-\nu} \left( 1 - \frac{1}{2} \frac{1}{\ln(\frac{4h}{w})} \right), \qquad (14)$$

181 $$\beta_\tau = \frac{\alpha}{4} \frac{1-2\nu}{1-\nu} \frac{1}{\ln(\frac{4h}{w})}. \qquad (15)$$

182 Using $\tau$ and, according to the Terzaghi's law (eq. 1), $\sigma'_m = \sigma_m - p$ in the *local* (i.e.

183 evaluated at the potential failure point) failure criteria allows predictions of failure

184 pattern and initiation criteria as a function of the *"global"* and undisturbed by the

185 localized fluid pressure rise far-field stresses $\sigma_V$ and $\sigma_H$. Equations 12-13 can be

186 interpreted as a generalized form of the Terzaghi's law expressed in terms of the far-
187 field stresses for the case of "local" fluid pressure not necessarily equal to the far-filed
188 fluid pressure. According to (eq. 15), during the fluid pressure increase, the radius of
189 the Mohr-circle decreases when $\frac{\sigma_V - \sigma_H}{2} - \frac{\eta}{2}\frac{p}{\ln(\frac{4h}{w})}$ is positive (patterns I and III) and
190 increases when it is negative (patterns II and IV). In the case when $\ln(\frac{4h}{w})$ 1, the
191 radius of the Mohr circle does not vary. If the rock is incompressible ($\nu = 0.5$) or if the
192 Biot-Willis coupling constant is set to zero, then equations (14-15) recovers the
193 expected Terzaghi's limit ($\beta_\sigma = 1$ and $\beta_\tau = 0$) Indeed, the case when the fluid pressure
194 gradients are not coupled to solid deformation must be in agreement with the classical
195 effective stress law well supported by experiments with a homogeneous fluid pressure
196 distribution [*Garg and Nur*, 1973; *Paterson and Wong*, 2005].
197 Thus, after evaluating initial stresses at depth $y = -h$ using (eq. 4) and substitution of $\tau$
198 and $\sigma'_m$ from (eqs. 12-13) using (eqs. 14-15) into the shear and tensile failure
199 conditions (eqs. 2-3), the critical pore-fluid pressure $p^c$ is calculated. Similarly, using
200 the analytical solution and equation (4) at the free surface we obtain $\sigma_H = 0$ and the
201 tensile failure condition: $2\beta_\tau p^c = \sigma_T$ for failure pattern V.
202 Based on the above calculations, we obtain a generalized expression for failure criteria
203 for all failure patterns as a linear combination of initial stresses evaluated at appropriate
204 depth:

205 $$k_b(\sigma_H - \sigma_V) = k_\tau + k_\sigma(\sigma_V + \sigma_H) - k_f p^c \qquad (16)$$

206  By rearranging, we obtain a unified expression for the critical pore-fluid pressure $p^c$:

207  $$p^c = \frac{k_b(\sigma_V - \sigma_H) + k_\tau + k_\sigma(\sigma_V + \sigma_H)}{k_f} \qquad (17)$$

208  where $k_b$, $k_\tau$, $k_\sigma$, and $k_f$ are constant coefficients (see Table 1) for the various failure

209  patterns shown on Figure 3. Using these coefficients and equation (17), the phase-

210  diagram for the different failure patterns can be calculated. The minimum value of $p^c$

211  for patterns I-V defines the pore-fluid overpressure at failure nucleation (Figure 4, red

212  colors). If $p^c \leq 0$, the rock is at failure without fluid overpressure (Figure 4, white

213  color). In Figure 4, the vertical axis corresponds to the vertical stress $\sigma_V$ and the

214  horizontal axis to the stress difference $(\sigma_V - \sigma_H)$, both axes being normalized by $C$,

215  the cohesion of the rock. Any initial stress state in the model corresponds to a point on

216  the diagram. The contours in the colored regions plot the dimensionless pressure

217  $p^{c*} = \frac{2\beta_\tau}{\sigma_T} p^c$ at failure onset. If the value of the localized pore-fluid pressure is smaller

218  than $p^{c*}$, then the system is stable. However, if it is equal or larger, then the porous

219  material fails with a predictable pattern-onset, that depends on the position in the phase

220  diagram.

221  White thick lines on Figure 4 define the topology of transition boundaries between

222  different failure patterns ($p^c_i = p^c_j$ where $i$ and $j$ are patterns I-V in (eq. 17) and

223  Table1, $i \neq j$ ). Their equations can be calculated using equation (17) and the

224  parameters given in Table 1. If a point in the failure diagram lies on one of these

225 transition lines, the yield functions for the two corresponding domains are both equal to
226 zero, implying that both failure modes could occur.

227 **5. Conclusion**

228 We present an analytical and numerical analysis of the effect of a localized pore-fluid
229 pressure source on the failure pattern of crustal rocks. The main results are the
230 following:
231 - Depending on the initial conditions, the geometry, and the material properties,
232 five different patterns of failure can be characterized, either with tensile or shear
233 mode.
234 - The critical fluid pressure at the onset of failure could also be determined for all
235 failure patterns and an analytical solution for $p^c$ is given in equation (17) and
236 Table 1.
237 These results can be used in many geological applications, including the formation of
238 hydrothermal vent structures triggered by sill intrusion [*Jamtveit et al.*, 2004], the
239 aftershocks activities caused by motions of fluids inside faults [*Miller et al.*, 2004], or
240 the tremors caused by sediments dehydration in subduction zones [*Shelly et al.*, 2006].
241 Finally, our simulations did not allow studying any transient effects in the fluid
242 pressure during fracture propagation. It has also been shown that fluid lubrication
243 [*Brodsky and Kanamori*, 2001] could have strong effect on the dynamics of rupture
244 propagation. We are currently neglecting these additional effects, which could be
245 integrated in an extended version of our model.
246

**Acknowledgments:** This work was financed by PGP (Physics of Geological Processes) a Center of Excellence at the University of Oslo.

**References**

Brodsky, E. E. and H. Kanamori (2001), The elastohydrodynamic lubrication of faults, *Journal of Geophysical Research*, 106(B8), 357-374.

Ingrafea, A. R. (1987), Theory of crack initiation and propagation in rock, in Fracture Mechanics of Rock, edited by B. K. Atkinson, pp. 71-110, Academic Press, London.

Garg, S. K., and A. Nur (1973), Effective stress laws for fluid-saturated porous rocks, *Journal of Geophysical Research*, 78(26), 5911-5921.

Jaeger, J. C. (1963), Extension Failures in Rocks Subject to Fluid Pressure, *Journal of Geophysical Research, 68*, 6066-6067.

Jamveit, B., H. Svensen, J.J. Podladchikov, and S. Planke (2004), Hydrothermal vent complexes associated with sill intrusions in sedimentary basins, *Geological Society of London, Special Publications*, *234*, 233-241.

Jones, A. (1971), Soil piping and stream channel initiation, *Water Resources Research, 7*, 602–610.

Miller S. A., C. Collettini, L. Chiaraluce, M. Cocco, M. Barchi, and B. Kaus (2004), Aftershocks driven by a high pressure $CO_2$ source at depth, *Nature, 427*, 724-727.

Mourgues R., and P.R. Cobbold (2003), Some tectonic consequences of fluid overpressures and seepage forces as demonstrated by sandbox modeling, *Tectonophysics, 376*, 75-97.

Murrell, S. A. F. (1964), The theory of propagation of elliptical Griffith cracks under various conditions of plane strain or plain stress. Parts 2, 3; *British Journal of Applied Physics, 15*, 1211-23.

Paterson, M. S., and Wong, T.-F. (2005), *Experimental Rock Deformation-The Brittle Field*, 346 pp., Springer, Berlin.

Poliakov, A.N.B, Y. Podladchikov, and C. Talbot (1993), Initiation of salt diapirs with frictional overburdens: numerical experiments, *Tectonophysics, 228*, 199-210.

Rice, J. R., and M. P. Cleary (1976), Some basic stress-diffusion solutions for fluid-saturated elastic porous media with compressible constituents, *Reviews of Geophysics and Space Physics, 14*, 227-241.

Shelly, D. R., G. C. Beroza, S. Ide, and S. Nakamula (2006), Low-frequency earthquakes in Shikoku, Japan and their relationship to episodic tremor and slip, *Nature, 442*, 188-191, doi:10.1038/nature04931.

Skempton, A.W. (1961), *Effective stress in soil, concrete and rocks,* in Pore Pressure and Suction in Soils, pp. 4-16, Butterworths, London.

Terzaghi, K. (1943), *Theoretical Soil Mechanics*, John Wiley and Sons, 528 pp., New York.

Vermeer, P.A. (1990), The orientation of shear bands in biaxial tests, *Géotechnique, 40*(2), 223-236.

289 **Figure captions**

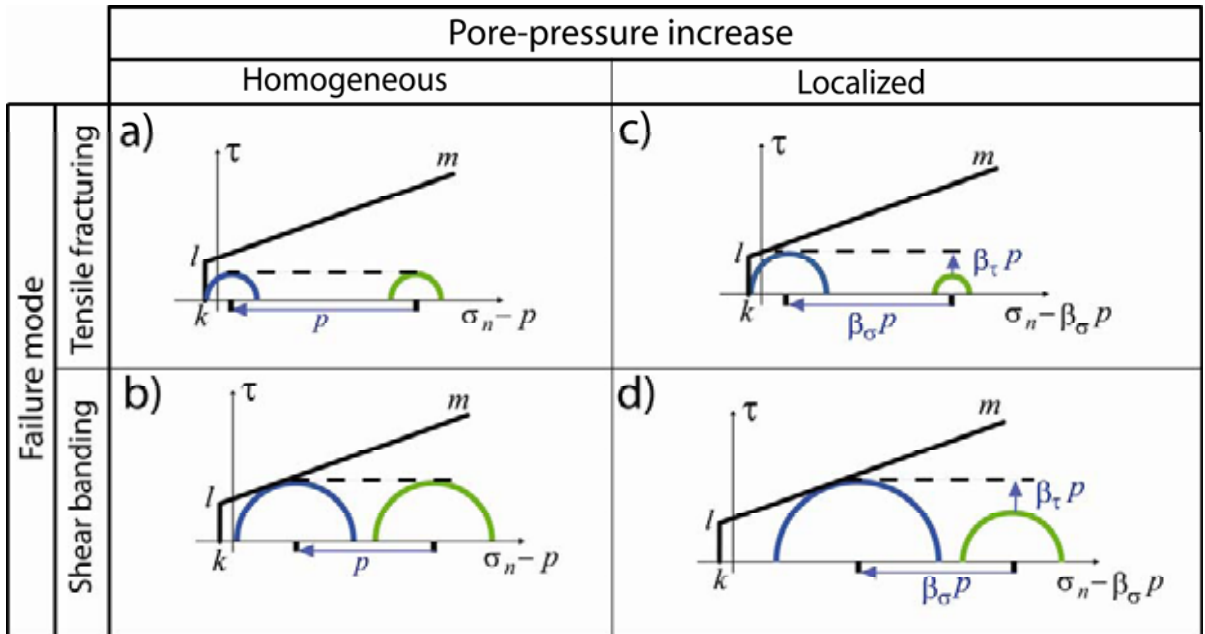

290
291 **Figure 1.** Effect of *homogeneous* (a, b) vs. *localized* (c, d) pore-fluid pressure increase
292 on the failure. $lm$ is the Mohr-Coulomb failure envelope (2); $kl$ is the tensile cut-off
293 boundary (3). The Mohr circles for initial stress conditions are represented in green.
294 The Mohr circles after a pore-fluid increase $p$ are represented in blue (at failure).
295 Arrows show the transformation of the Mohr circle after pore-fluid pressure increase.
296 The rock fails either in shear mode, or in tension, when the Mohr circle meets the
297 failure envelopes $kl$ or $lm$, respectively. Note that for localized pore-fluid pressure
298 increase, the radius of the Mohr circle is changed (increase or decrease) by an amount
299 $\beta_\tau p$ and the center is displaced to the left by an amount $\beta_\sigma p$. The two dimensionless
300 parameters $\beta_\tau$ and $\beta_\sigma$ are given in equations 14-15.

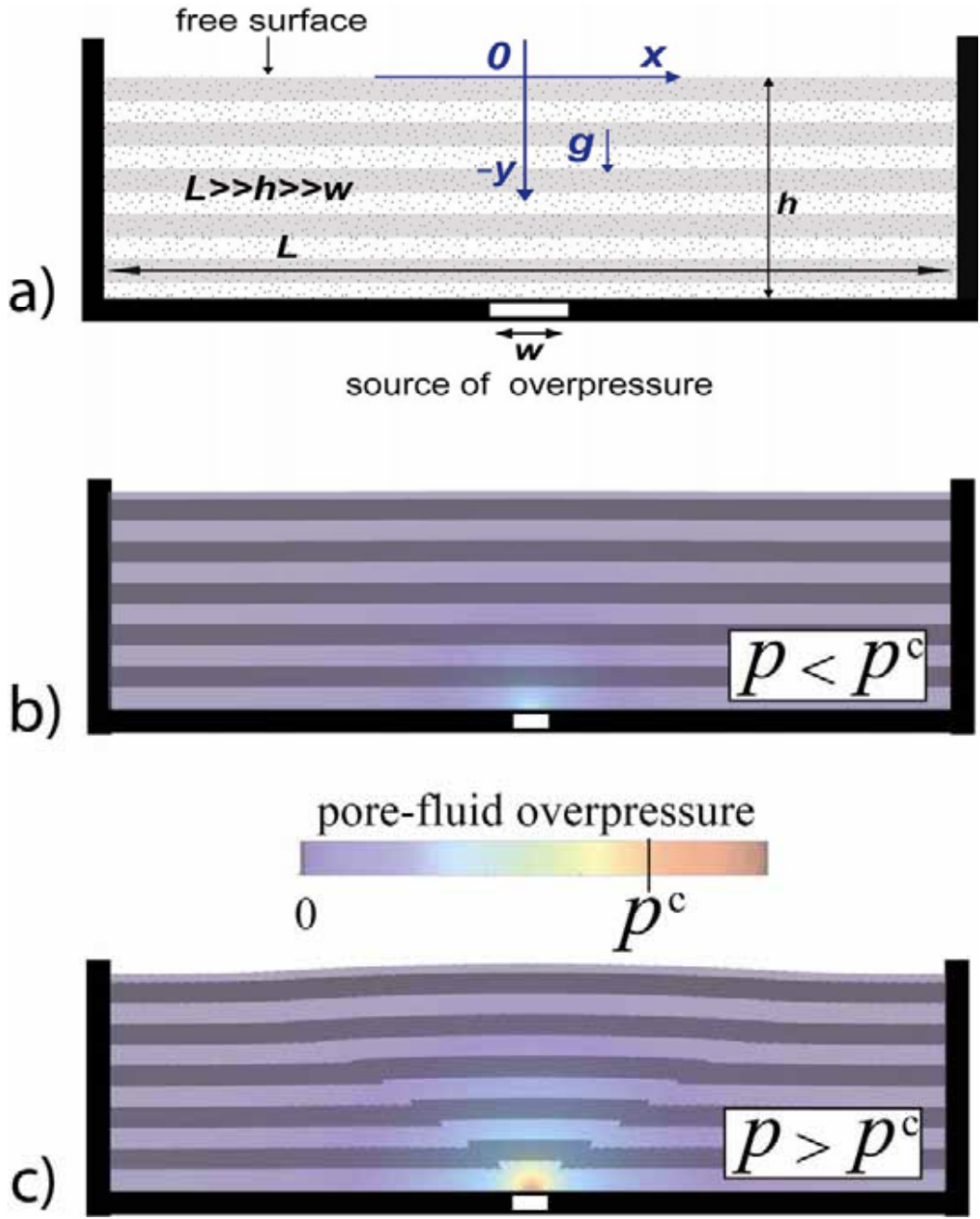

**Figure 2.** Geometry of the plane-strain model (a) and representative stages of deformation before (b) and during failure (c). The color coding represents the pore-fluid pressure normalized to the pressure at the onset of failure $p^c$. The horizontal layering represents passive markers of the deformation.

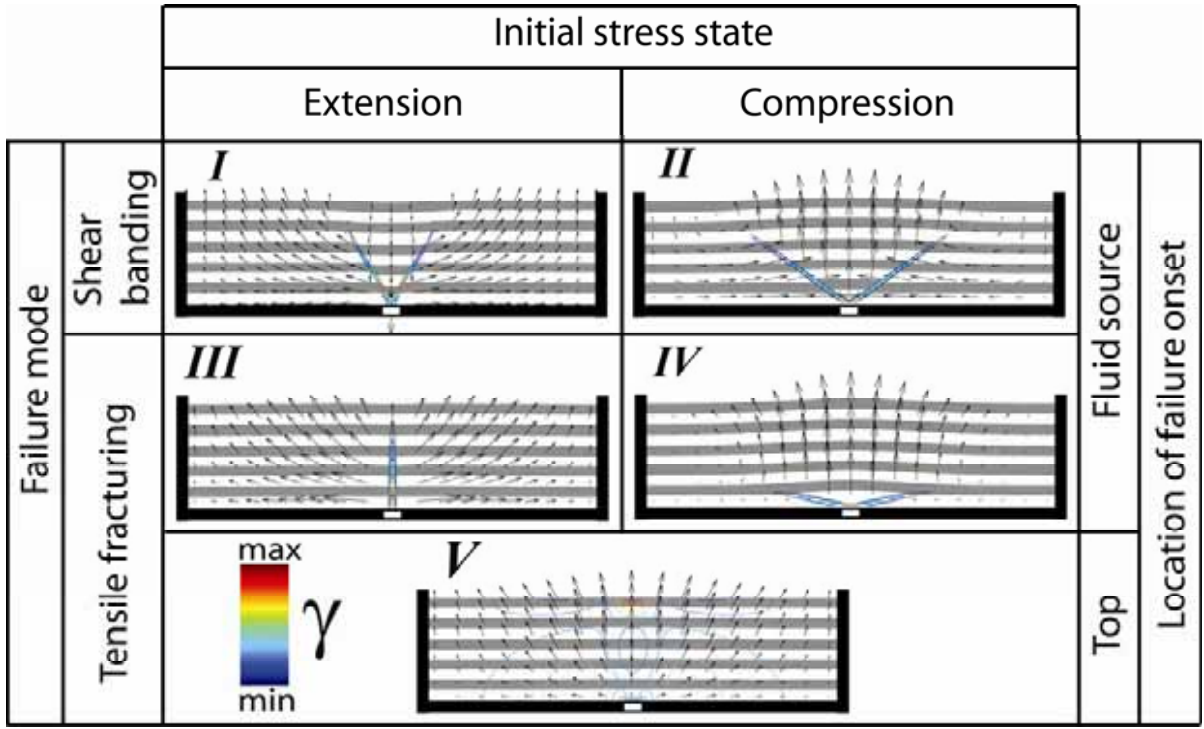

**Figure 3.** Possible failure patterns caused by the localized pore-fluid pressure increase. Either shear bands (I, II) or tensile fractures (III, IV, V) can develop. Fracture nucleates either on the fluid source at depth (I, II, III, IV) or on the free surface (V). Horizontal passive marker layering, arrows, and color coding (contours of strain deviator $\gamma$) indicate the displacement and intensity of the deformation. (See Auxiliary Materials for animations and for numerical parameters used in simulations).

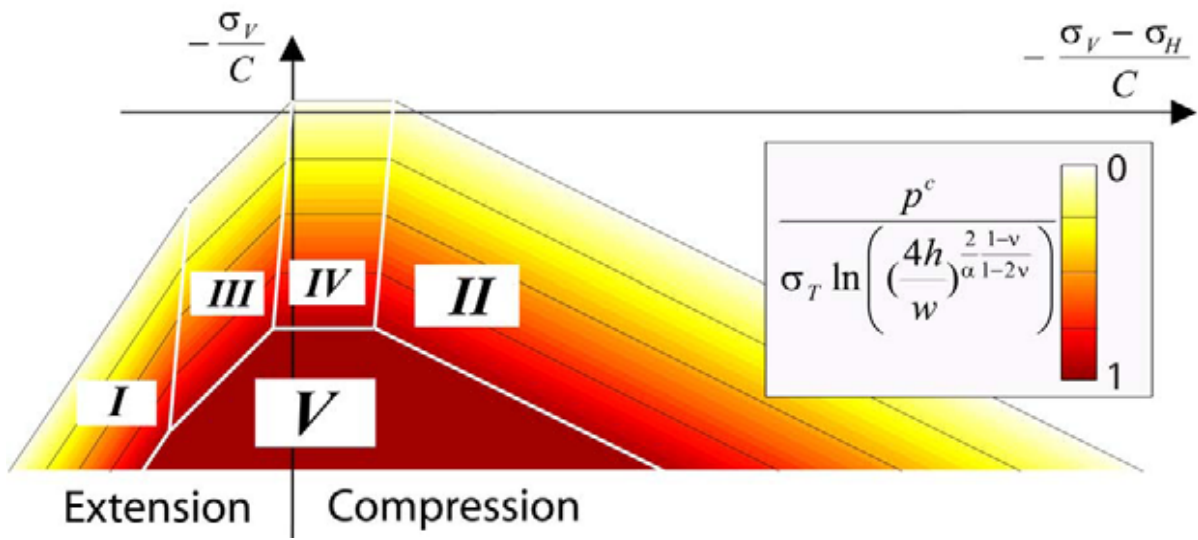

**Figure 4.** Phase diagram of failure-onset patterns. On the vertical axis the non-dimensional vertical stress is plotted, while on the horizontal axis the non-dimensional stress-difference between vertical and horizontal stresses is plotted. Stresses are given at the fluid source at depth. The colored region represents the admissible stress state at which the rock is stable, while the outer region represents the unstable combination of stresses. The colors plot the non-dimensional critical fluid pressure. The bold white lines represent the topology of transitions between the different failure modes. Interestingly, failure patterns IV and V may occur both in extensional and in compressive initial stress state.

325 **Table 1.** Critical pore-pressure $p^c = \dfrac{k_b(\sigma_V - \sigma_H) + k_\tau + k_\sigma(\sigma_V + \sigma_H)}{k_f}$ at the onset of
326 failure, corresponding to the various failure patterns (I-V) shown on Figure 3. The
327 coefficients also define the domains in the phase diagram of Figure 4.
328

|   | $k_f$ | $k_\tau$ | $k_\sigma$ | $k_b$ |
|---|---|---|---|---|
| I | $2(\beta_\tau + \sin(\phi)\beta_\sigma)$ | $2C\cos(\phi)$ | $\sin(\phi)$ | 1 |
| II | $2(\beta_\tau - \sin(\phi)\beta_\sigma)$ | $-2C\cos(\phi)$ | $-\sin(\phi)$ | 1 |
| III | $2(\beta_\tau + \beta_\sigma)$ | $2\sigma_T$ | 1 | 1 |
| IV | $2(\beta_\tau - \beta_\sigma)$ | $-2\sigma_T$ | -1 | 1 |
| V | $2\beta_\tau$ | $\sigma_T$ | 0 | 0 |

329

**Auxiliary Materials:**

**Failure patterns caused by localized rise in pore-fluid overpressure and effective strength of rocks**

A.Y. Rozhko[1], Y.Y. Podladchikov[1], and F. Renard[1,2]

[1] Physics of Geological Processes, University of Oslo, PO box 1048 Blindern, 0316 Oslo, Norway

[2] LGCA-CNRS-OSUG, University of Grenoble, BP 53, 38041 Grenoble, France

In Section A of the Auxiliary Materials, we demonstrate the analytical solution and derive equations (12-17) of Section 4 in the article. Section B is devoted to the comparison of the analytical solution to the numerical tests. Finally, some animations of the results shown in Figure 3 are presented in Section C.

**A. Analytical Solution**

In this section we calculate the seepage forces caused by coupling pore-fluid diffusion with rock deformation. The procedure for calculating these forces can be divided into six successive steps:

1) The definition of the system of equations for mechanical equilibrium;

2) the general solution of this system of equations in Cartesian coordinates using the complex potential method for poro-elasticity;

3) the introduction of the curvilinear coordinate system associated with the conformal mapping transformation, which allows finding the solution for complex geometry;

4) the calculation of the general solution of the equilibrium equations for steady-state poro-elasticity in curvilinear coordinate system;

5, 6) and finally, after defining the boundary conditions, the calculation of the particular solution.

**A1. System of equations for steady-state poro-elastic deformation**

Following the common approach of [*Biot*, 1941; *Rice and Cleary*, 1976], the equations for steady-state fluid filtration in a poro-elastic solid are given by:

- The stress balance equations for the total stress tensor $\sigma_{ij}$, without volume forces:

$$\frac{\partial \sigma_{xx}}{\partial x} + \frac{\partial \sigma_{xy}}{\partial y} = 0 \text{ and } \frac{\partial \sigma_{yy}}{\partial y} + \frac{\partial \sigma_{xy}}{\partial x} = 0 \tag{S1}$$

- The steady-state fluid filtration governed by the Laplace equation for fluid pressure $p^f$:

$$\left(\frac{\partial^2}{\partial x^2} + \frac{\partial^2}{\partial y^2}\right) p^f = 0 \tag{S2}$$

- A constitutive relation between the total stress $\sigma_{ij}$ and the strain $\varepsilon_{ij}$:

$$\varepsilon_{ij} = \frac{\sigma_{ij} - \sigma_m \delta_{ij}}{2G} + \frac{\sigma_m - \alpha p^f}{2G} \delta_{ij} \frac{1-2\nu}{1+\nu} \tag{S3}$$

where $\delta_{ij}$ is the Kronecker delta and the intrinsic material properties are the shear modulus $G$, the drained Poisson's ratio $\nu$, and the Biot-Willis poro-elastic constant $\alpha$ [*Paterson and Wong*, 2005].

**A2. Complex Potential Method for steady-state poro-elasticity**

Two-dimensional problems in elasticity can be solved using a complex potential method (CPM), developed by Kolosov [1909] and Muskhelishvili [1977]. This method has been generalized for thermo-elasticity by Lebedev [1937] and others. In this method the general solution for the displacements and stresses is represented in terms of two analytical functions (potentials) of a complex variable and another complex function for temperature distribution [*Goodier and Hodge*, 1958; *Timoshenko and Goodier*, 1982]. This general solution automatically satisfies the force balance equation and generalized Hooke's law for thermo-elasticity, provided that the function of temperature distribution satisfies to the heat conduction equation. The nontrivial part of this method is in satisfying the boundary conditions of the specific problem, which is done by conformal mapping.

The equations for poro-elasticity and thermo-elasticity are identical for steady-state fluid filtration and heat flow problems. Therefore it is possible to use the complex potential method, developed for thermo-elasticity, to solve steady-state poro-elastic problems.

According to the CPM, the general solution of equations (S1-S3) can be written in the form (by convention, compressive stress and strain are positive) [*Goodier and Hodge*, 1958; *Timoshenko and Goodier*, 1982; *Muskhelishvili*, 1977]:

$$\sigma_{xx} + \sigma_{yy} = -4\operatorname{Re}[\varphi'(z)] + 2\eta p^f(z,\bar{z}) \tag{S4}$$

$$\sigma_{yy} - \sigma_{xx} + 2i\sigma_{xy} = -2[\bar{z}\varphi''(z) + \psi'(z)] + 2\eta \int \frac{\partial p^f(z,\bar{z})}{\partial z} d\bar{z} \tag{S5}$$

$$2G(u_x + iu_y) = \chi\varphi(z) - z\overline{\varphi'(z)} - \overline{\psi(z)} + \eta \int p^f(z,\bar{z}) dz \tag{S6}$$

Where the plane-strain parameters $\eta$ and $\chi$ are defined as: $\eta = \frac{\alpha}{2}\frac{1-2\nu}{1-\nu}$ and $\chi = 3 - 4\nu$. The integrals in (eqs. S5-S6) are indefinite (without the integration constant) and the superscript "'" denotes differentiation, i.e. $\varphi'(z) = \frac{\partial \varphi(z)}{\partial z}$.

In equations (S4-S6), $z = x + iy$ is the complex variable, $x$ and $y$ are the usual Cartesian coordinates, $i = \sqrt{-1}$ is the imaginary unit, and the overbar denotes complex conjugation, i.e. $\bar{z} = x - iy$. $\varphi(z)$ and $\psi(z)$ are the complex potentials, which are analytic functions of the complex variable $z$, and are derived from the biharmonic Airy function [*Muskhelishvili*, 1977]. $p^f(z,\bar{z})$ is the solution of Laplace equation (S2), given as a function of two complex variables $z$ and $\bar{z}$. By introducing the transformation of coordinates $x = \frac{z+\bar{z}}{2}$ and $y = \frac{z-\bar{z}}{2i}$, the Laplace equation (S2) can be rewritten in the form [*Timoshenko and Godier*, 1982; *Lavrent'ev and Shabat*, 1972]:

$$\frac{\partial^2}{\partial z \partial \bar{z}} p^f(z,\bar{z}) = 0. \tag{S7}$$

The fluid filtration pressure creates stresses at the boundaries of the solid. The boundary value problem, with given stresses or displacements in curvilinear boundaries, can be solved by finding the complex potentials $\varphi(z)$ and $\psi(z)$ using Muskhelishvili's method.

**A3. Conformal transformation and curvilinear coordinates**

Conformal mapping is a transformation of coordinates that allows for solving a problem with a simple geometry (Figure S1a) and transforming its solution to a more complex geometry (Figure S1b). The properties of conformal mapping can be found in [*Lavrent'ev and Shabat*, 1972].

The transformation of a circular domain into an elliptical one is given in a unique way by the Joukowsky transform:

$$z = \frac{w}{4}(\varsigma + \frac{1}{\varsigma}) \qquad (S8)$$

where $w$ is the width of pore-fluid pressure source as defined in Figure 2 (and Figure S1b, red line). The complex variable $\varsigma$ is defined through polar coordinates $\rho$ and $\vartheta$ as follows (Figure S1a):

$$\varsigma = \rho e^{i\vartheta}. \qquad (S9)$$

In Figure S1a, $1 \le \rho \le \rho_*$ and $0 \le \vartheta \le 2\pi$; the internal ($\rho = 1$) and external ($\rho = \rho_*$) boundaries are shown by the red and brown curves, respectively.

We use the polar coordinates $\rho$ and $\vartheta$ in the $\varsigma$-plane, as a non-dimensional system of coordinates for the $z$-plane. The properties of this coordinate system are considered below. Any circle $\rho = const$ and radius $\vartheta = const$ in the $\varsigma$-plane (Figure S1a) are transformed into an ellipse and a hyperbola in the z-plane, respectively (Figure S1b). The foci of the ellipse and the hyperbola on the z-plane coincide (Figure S1b). As the conformal mapping preserves angles, the two lines $\vartheta = const$ and $\rho = const$ are perpendicular in the z-plane. Therefore, the polar coordinates $\rho$ and $\vartheta$ can be considered as a curvilinear coordinate system in the $z$-plane.

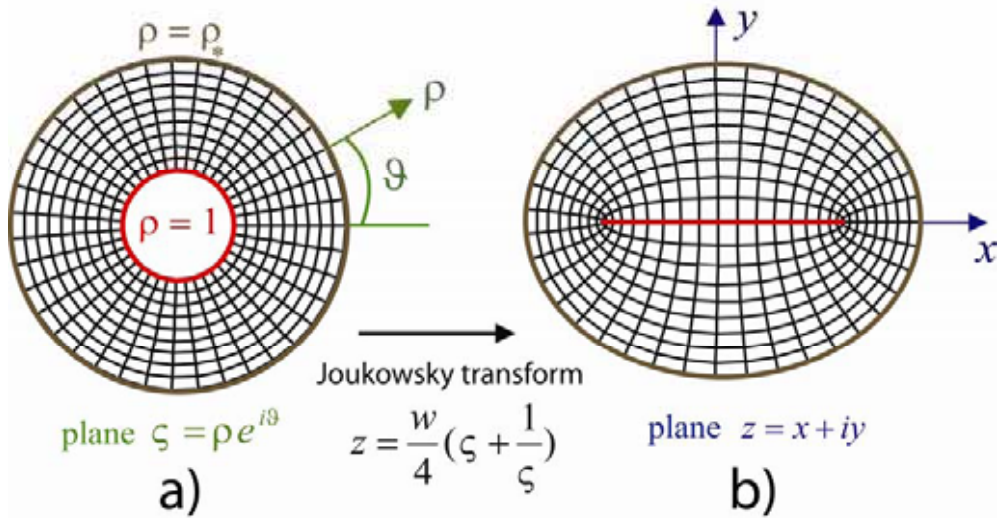

**Figure S1.** Conformal mapping procedure using the Joukowsky transform and systems of Cartesian and curvilinear coordinates. Here $w$ is the length of pore pressure source (red line segment) shown on Figure 2a.

The Cartesian and curvilinear coordinates on the z-plane are related through:

$$x = \frac{w}{4}\rho(1+\frac{1}{\rho^2})\cos(\vartheta)$$
$$y = \frac{w}{4}\rho(1-\frac{1}{\rho^2})\sin(\vartheta)$$
(S10)

The Cartesian coordinates of the pore-pressure source (Figure S1b, red line) are given by substitution of the circle $\rho = 1$ (Figure S1a, red circle) into (eq. S10):

$$x = \frac{w}{2}\cos(\vartheta)$$
$$y = 0$$
(S11)

The external boundaries $\rho = \rho_*$ are represented as brown curves on Figure S1. If $\rho_* \gg 1$, the relations (S10) become:

$$x = \frac{w}{4}\rho_* \cos(\vartheta) + O\left(\frac{1}{\rho_*^2}\right)$$
$$y = \frac{w}{4}\rho_* \sin(\vartheta) + O\left(\frac{1}{\rho_*^2}\right) \quad \text{for } \rho = \rho_* \gg 1$$
(S12)

By neglecting the terms $O\left(\frac{1}{\rho_*^2}\right)$ and taking $\rho_* = \frac{4h}{w}$, the relationship (S12) becomes

$$x = h\cos(\vartheta)$$
$$y = h\sin(\vartheta)$$
(S13)

These equations describe the external boundary, which is given by a circle with radius $h$, for the case when $\frac{4h}{w} \gg 1$ (Figure S1b, brown curve). We derive the analytical solution for this case where $\frac{4h}{w} \gg 1$.

### A4. General solution of plane poro-elasticity in curvilinear coordinates

According to Muskhelishvili [1977], the stress and the displacement components in the Cartesian and Curvilinear coordinate systems are related by:

$$\sigma_{xx} + \sigma_{yy} = \sigma_{\rho\rho} + \sigma_{\vartheta\vartheta} \tag{S14}$$

$$\sigma_{yy} - \sigma_{xx} + 2i\sigma_{xy} = \frac{\rho^2}{\varsigma^2}\frac{\overline{\omega'(\varsigma)}}{\omega'(\varsigma)}\left(\sigma_{\vartheta\vartheta} - \sigma_{\rho\rho} + 2i\sigma_{\rho\vartheta}\right) \tag{S15}$$

and

$$u_x + iu_y = \frac{\rho}{\varsigma} \frac{\overline{\omega'(\varsigma)}}{\omega'(\varsigma)} (u_\rho + iu_\vartheta) \tag{S16}$$

where

$$\omega(\varsigma) = z = \frac{w}{4}(\varsigma + \frac{1}{\varsigma}) \tag{S17}$$

Note that, if $\rho = 1$ in equations (S15-S16), one obtains using (eq. S17):

$$\frac{\rho^2}{\varsigma^2} \frac{\overline{\omega'(\varsigma)}}{\omega'(\varsigma)} = -1 \text{ and } \frac{\rho}{\varsigma} \frac{|\omega'(\varsigma)|}{\overline{\omega'(\varsigma)}} = i \frac{\sqrt{\sin^2(\vartheta)}}{\sin(\vartheta)} = \begin{cases} i \text{ for } 0<\vartheta<\pi \\ -i \text{ for } \pi<\vartheta<2\pi \end{cases} \tag{S18}$$

The equations for stresses (S4-S5) and displacements (S6) in the curvilinear coordinate system become:

$$\sigma_{\rho\rho} + \sigma_{\vartheta\vartheta} = -4\operatorname{Re}(\Phi(\varsigma)) + 2\eta p^f(\varsigma,\overline{\varsigma}) \tag{S19}$$

$$\sigma_{\vartheta\vartheta} - \sigma_{\rho\rho} + 2i\sigma_{\rho\vartheta} = -\frac{\varsigma^2}{\rho^2} \frac{2}{\omega'(\varsigma)}\left[\overline{\omega(\varsigma)}\Phi'(\varsigma) + \omega'(\varsigma)\Psi(\varsigma)\right] + \frac{\varsigma^2}{\rho^2} \frac{2}{\omega'(\varsigma)} \eta \int \frac{\partial p^f(\varsigma,\overline{\varsigma})}{\partial \varsigma} \overline{\omega'(\varsigma)} d\overline{\varsigma} \tag{S20}$$

$$2G(u_\rho + iu_\vartheta) = \frac{\overline{\varsigma}}{\rho} \frac{\overline{\omega'(\varsigma)}}{|\omega'(\varsigma)|} \left( \chi\varphi(\varsigma) - \frac{\omega(\varsigma)}{\overline{\omega'(\varsigma)}}\overline{\varphi'(\varsigma)} - \overline{\psi(\varsigma)} + \eta \int p^f(\varsigma,\overline{\varsigma})\omega'(\varsigma)d\varsigma \right) \tag{S21}$$

where

$$\Phi(\varsigma) = \frac{\varphi'(\varsigma)}{\omega'(\varsigma)} \text{ and } \Psi(\varsigma) = \frac{\psi'(\varsigma)}{\omega'(\varsigma)} \tag{S22}$$

Using the properties of conformal mapping ($\omega'(\varsigma) \neq 0$ and $\omega(\varsigma) \neq 0$), the Laplace equation (S7) becomes:

$$\frac{\partial^2 p^f(\varsigma,\overline{\varsigma})}{\partial \varsigma \partial \overline{\varsigma}} = 0 \tag{S23}$$

### A5. Boundary conditions

The boundary conditions for the fluid are:

$$\begin{aligned} p^f &= p \text{ for } \rho = 1 \\ p^f &= 0 \text{ for } \rho = \rho_* \end{aligned} \tag{S24}$$

and the boundary conditions for the solid are:

$$\sigma_{xy} = 0 \text{ and } u_y = 0 \text{ for } \rho = 1$$
$$\sigma_{\rho\vartheta} = 0 \text{ and } \sigma_{\rho\rho} = 0 \text{ for } \rho = \rho_*$$
(S25)

Note here that if $\rho = 1$, and using (eqs. S14-S18), one obtains $\sigma_{\rho\vartheta} = -\sigma_{xy} = 0$ and $u_\rho = \pm u_y = 0$.

We define the pore-fluid pressure source inside the continuous medium (Figure S1b, red line) as a line segment with constant fluid pressure. The external boundary (Figure S1b, brown line) has zero pore pressure. We use mixed boundary conditions for the pore pressure source $\rho = 1$ as in (eq. S25) because the medium is continuous everywhere and $y = 0$ is a symmetry line. The external boundary $\rho = \rho_*$ is free from load.

## A6. Solution

We finally present the analytical solution derived using Muskhelishvili's method. We do not show the derivation here, since it is quite lengthy, but we demonstrate that our solution fulfills to the boundary conditions.

The solution of Laplace equation (S23) with the boundary conditions (S24) is given by,

$$p^f(\varsigma,\bar{\varsigma}) = p - p \frac{\ln(\varsigma\bar{\varsigma})}{\ln(\rho_*^2)}$$
(S26)

This equation can be simplified, using (eq. S8) as follows

$$p^f(\rho) = p - p \frac{\ln(\rho)}{\ln(\rho_*)}$$
(S27)

The boundary conditions (S24) are fulfilled by (eq. S27). This equation (S27) gives the solution for pore fluid pressure.

We calculate the complex potentials, which define the solution of problem, as the following:

$$\varphi(\varsigma) = \frac{p\,\eta\,w}{16} \frac{\varsigma^2 + 1}{\varsigma \ln(\rho_*)}$$
(S28)

$$\psi(\varsigma) = \frac{p\,\eta\,w}{4\varsigma \ln(\rho_*)}$$
(S29)

The explicit expression for the stress components can be found after substitution of (eq. S26), and (eqs. S28-S29) into (eqs. S19-S20) using (eq. S22), and after simplifications:

$$\sigma_{\rho\rho} = -\eta \frac{p}{\ln(\rho_*)} \left( \ln\left(\frac{\rho}{\rho_*}\right) + 1 - \frac{(\rho^2 - 1)(\rho^2 - \cos(2\vartheta))}{1 - 2\rho^2 \cos(2\vartheta) + \rho^4} \right) \quad (S30)$$

$$\sigma_{\vartheta\vartheta} = -\eta \frac{p}{\ln(\rho_*)} \left( \ln\left(\frac{\rho}{\rho_*}\right) + 1 + \frac{(\rho^2 + 1)(\cos(2\vartheta) - 1)}{1 - 2\rho^2 \cos(2\vartheta) + \rho^4} \right) \quad (S31)$$

$$\sigma_{\rho\vartheta} = -\eta \frac{p}{\ln(\rho_*)} \frac{(\rho^2 - 1)\sin(2\vartheta)}{1 - 2\rho^2 \cos(2\vartheta) + \rho^4} \quad (S32)$$

The explicit expression for the displacements can be found after substitution of (eqs. S26, S28-S29) into (eq. S21), using (eqs. S16, S22), and after simplifications we obtain:

$$u_x = w \frac{\eta}{32} \frac{p \cos(\vartheta)}{G \ln(\rho_*)} \left[ \left( 4\ln\left(\frac{\rho_*}{\rho}\right) + \chi + 1 \right)\left(\rho - \frac{1}{\rho}\right) + \left( 4\ln\left(\frac{\rho_*}{\rho}\right) + \chi - 3 \right)\frac{2}{\rho} \right]$$

(S33)

$$u_y = w \frac{\eta}{32} \frac{p \sin(\vartheta)}{G \ln(\rho_*)} \left( 4\ln\left(\frac{\rho_*}{\rho}\right) + \chi + 1 \right)\left(\rho - \frac{1}{\rho}\right) \quad (S34)$$

In equations (S33-S34), we present the analytical solution for the displacements along $x$ and $y$ axes. This solution is parameterized through curvilinear coordinates $\rho$ and $\vartheta$. The solution for displacements along $\rho$ and $\vartheta$ axes is too long to be reproduced here. One can find this displacements using (eqs. S33-S34) along with (eq. S16).

Applying the boundary conditions (S25) to equations (S32, S34) shows that the solution is fulfilled at the pore overpressure source $\rho = 1$ (Figure S1, red curve). The non-zero stress components at the pore overpressure source are calculated using (eqs. S30-S31) along with (eqs. S14-S15) and (eq. S18):

$$\sigma_{yy} = \sigma_{\rho\rho} = \eta p - \frac{\eta p}{\ln(\rho_*)} \text{ for } \rho = 1 \quad (S35)$$

$$\sigma_{xx} = \sigma_{\vartheta\vartheta} = \eta p \text{ for } \rho = 1 \quad (S36)$$

According to equation (1) of the article and equations (S27) and (S35-S36), the effective stress at the pore overpressure source becomes

$$\sigma'_{yy} = \eta p - \frac{\eta p}{\ln(\rho_*)} - p \text{ for } \rho = 1 \quad (S37)$$

$$\sigma'_{xx} = \eta p - p \text{ for } \rho = 1 \tag{S38}$$

At the external boundary $\rho = \rho_*$ (Figure S1, brown curve), we obtain $\sigma_{\rho\rho} = O\left(\dfrac{1}{\rho_*^2}\right)$ and $\sigma_{\rho\vartheta} = O\left(\dfrac{1}{\rho_*^2}\right)$, since $\rho_* \gg 1$. The boundary conditions (S25) at $\rho = \rho_*$ are also fulfilled.

The circumferential stress at the free surface is given by:

$$\sigma_{\vartheta\vartheta} = -\eta \frac{p}{\ln(\rho_*)} + O\left(\frac{1}{\rho_*^2}\right) \quad (\sigma_{\vartheta\vartheta} = \sigma_{xx} \text{ for } \rho = \rho_* \text{ and } \vartheta = \frac{\pi}{2}) \tag{S39}$$

According to equation (1) of the paper and equations (S27) and (S39), the non-zero component of effective stress at the free surface is given by

$$\sigma'_{xx} = -\eta \frac{p}{\ln(\rho_*)} \tag{S40}$$

**A7. Critical pore pressure**

We now propose an analytical study in order to derive equation (17) of the paper for the fluid pressure $p^c$ at the onset of failure. To do this, we add the initial state of stress given by equations (4) to the stress of state which is exerted by the fluid overpressure increase (obtained in the Section A6). This is possible due to the additivity of linear poroelasticity [*Rice and Cleary*, 1976]. Both the numerical simulations and the analytical solution show that failure initiation takes place either at the fluid source or at the free surface. Therefore, in order to calculate $p^c$, we consider the stress and failure conditions below, first at the fluid source (patterns I-IV in Figure 3), and second at the free surface (pattern V in Figure 3).

Using equations (S37-S38) at ($\rho_* = \dfrac{4h}{w}$) and (eq. 4) (at depth $y = -h$) for failure patterns I-IV, the Terzaghi's effective stress tensor at the fluid source segment becomes:

$$\sigma'_{xx} = \sigma_H - p + \eta p, \tag{S41}$$

$$\sigma'_{yy} = \sigma_V - p + \eta p - \frac{\eta p}{\ln(\frac{4h}{w})}, \tag{S42}$$

$$\sigma'_{xy} = 0. \tag{S43}$$

The analytical solution given in equations (S41-S45) indicates that the stress components on the fluid source do not depend on *x* (Figure 2a).

Using (eq. S41-S45), the mean Terzaghi's effective stress $\sigma'_m$ and the stress deviator $\tau$ can be calculated:

$$\sigma'_m = \frac{\sigma_H + \sigma_V}{2} - p + \eta p - \frac{\eta}{2} \frac{p}{\ln(\frac{4h}{w})} \tag{S44}$$

$$\tau = \left| \frac{\sigma_V - \sigma_H}{2} - \frac{\eta}{2} \frac{p}{\ln(\frac{4h}{w})} \right|. \tag{S45}$$

Substitution of equations (S44) and (S45) into the failure condition (2) for shear failure or condition (3) for tensile failure defines $p^c$ for failure patterns 1-IV. Now, referring to the parameters $\beta_\sigma$ and $\beta_\tau$ defined in Figure 1 and to the definition of $\eta$ after equation (S6), one obtains from (eqs. S44, S45):

$$\beta_\sigma = 1 - \frac{\alpha}{2} \frac{1-2\nu}{1-\nu} \left( 1 - \frac{1}{2} \frac{1}{\ln(\frac{4h}{w})} \right), \tag{S46}$$

$$\beta_\tau = \frac{\alpha}{4} \frac{1-2\nu}{1-\nu} \frac{1}{\ln(\frac{4h}{w})}. \tag{S47}$$

According to (eq. S47), during the fluid pressure increase, the radius of the Mohr-circle decreases when $\frac{\sigma_V - \sigma_H}{2} - \frac{\eta}{2} \frac{p}{\ln(\frac{4h}{w})}$ is positive (Patterns I and III) and increases when it is negative (Patterns II and IV) in the case, if $\ln(\frac{4h}{w}) \gg 1$ the radius of Mohr circle does not change. If the rock is incompressible ($\nu = 0.5$) or the Biot-Willis coupling constant is set to zero equations (eqs. S46-S46) give $\beta_\sigma = 1$ and $\beta_\tau = 0$ therefore the seepage force does not have an additional effect on failure in this case.

For pattern V in Figure 3, the initial stresses (eq. 4) are zero (at depth $y = 0$). Thus after substitution of (eq. S40) into (eq. 3) and simplification we obtain the condition for tensile failure at the free surface as the following,

$$2\beta_\tau p^c = \sigma_T. \tag{S48}$$

By assuming $\dfrac{2C}{1+\sin(\varphi)} > \sigma_T$ we obtain that only tensile failure initiation is allowed at the free surface.

### B. Analytical versus numerical approach

The numerical simulations show that if the initial state of stress ($\sigma_V$ and $\sigma_H$) is taken as in the form of equations (4), then the nucleation of failure is allowed either at the pore-pressure source or at the free surface. We compare $p^c$ predicted with the analytical solution ((eq. 17) and Table 1) obtained *for the geometry shown on Figure S2a,* with $p^c$ calculated with the finite element method *for the geometry shown on Figure S2b*.

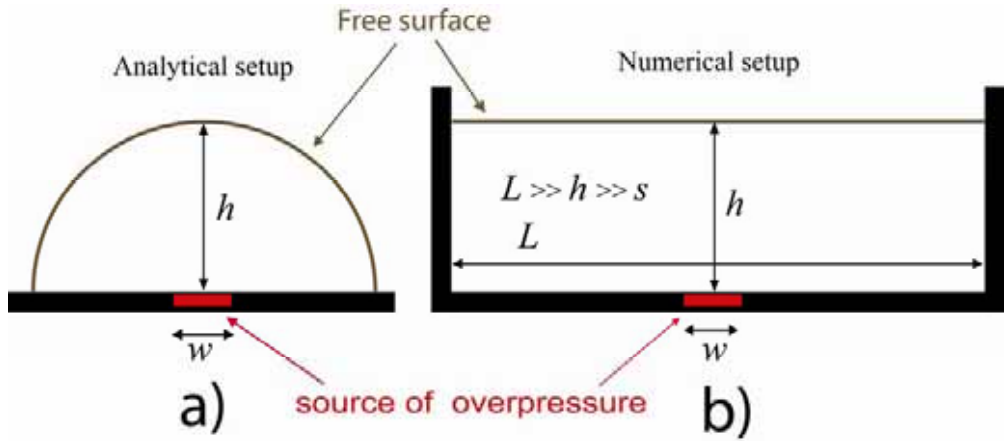

**Figure S2.** Geometry used in the analytical solution (Figure S1) compared to the geometry used in the numerical model (Figure 2a).

We studied $p^c$ numerically as a function of all parameters of the model. The vertical stress $\sigma_V$ (Pa or bar) and the depth $h$ (m) are chosen to be equal to 1 and all the parameters listed below are non-dimensional compared to these values. Plots on Figure S3 compare $p^c$ predicted with the analytical solution ($p^c_{AS}$) to $p^c$ calculated with the numerical simulations ($p^c_{NS}$); the different colors representing the various failure patterns. The value of $p^c$ is always within a 20% limit between the two approaches.

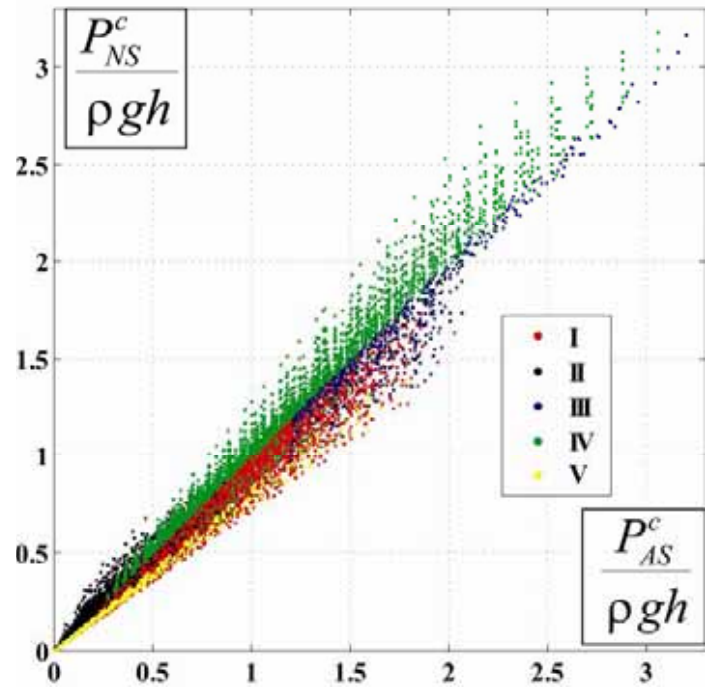

**Figure S3.** Critical pore pressure at the onset of failure predicted using the analytical solution $p^c_{AS}$ for the geometry shown on Figure S2a compared to the solution of the numerical simulations $p^c_{NS}$ for the geometry shown on Figure S2b. Each point corresponds to a single simulation. The different colors correspond to the different failure patterns of Figure 3

**List of parameters and values:**

Poisson's ratio: $\nu = \frac{[1:4]}{10}$ (where the notation $[1:4]$ denotes the array $[1,2,3,4]$).

Non-dimensional width of overpressure source: $\frac{w}{h} = \frac{2^{-[0:5]}}{5}$.

Non-dimensional horizontal stress: $\frac{\sigma_H}{\sigma_V} = \frac{[1:30]}{10}$.

Non-dimensional cohesion: $\frac{C}{\sigma_V} = e^{\ln(1e-4)+(\ln(3)-\ln(1e-4))\frac{[0:50]}{50}}$.

Non-dimensional tensile strength: $\frac{\sigma_T}{\sigma_V} = \frac{C}{\sigma_V}\left[\frac{1}{3}, \frac{1}{7}, \frac{1}{30}\right]$.

Friction angle (in degrees): $\varphi = 10^o[1:3]$.

Fixed non-dimensional parameters: $\dfrac{L}{h}=4$, $\dfrac{G}{\sigma_V}=10^7$, $\upsilon=0^o$ and $\alpha=1$.

## C. Animations of the simulations of Figure 3

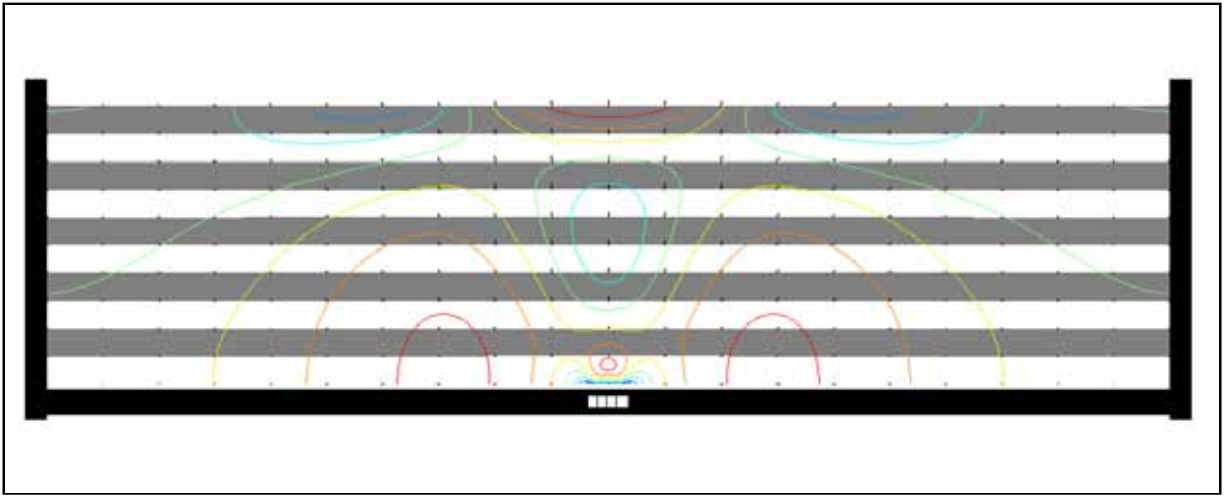

**Pattern I:**

$\dfrac{\sigma_H}{\sigma_V}=0.2$, $\varphi=33^o$, $\upsilon=0^o$, $\dfrac{C}{\sigma_V}=0.1$, $\dfrac{\sigma_T}{\sigma_V}=1$, $\nu=0.3$, $\dfrac{G}{\sigma_V}=10^7$, $\dfrac{L}{h}=4$, $\dfrac{w}{h}=10^{-1}$, $\alpha=1$. The first frame in the animation corresponds to the elastic solution at failure onset $p=p^c$, the last frame corresponds to the case when $p=3.5p^c$. The pore pressure increases linearly with time during the animation.

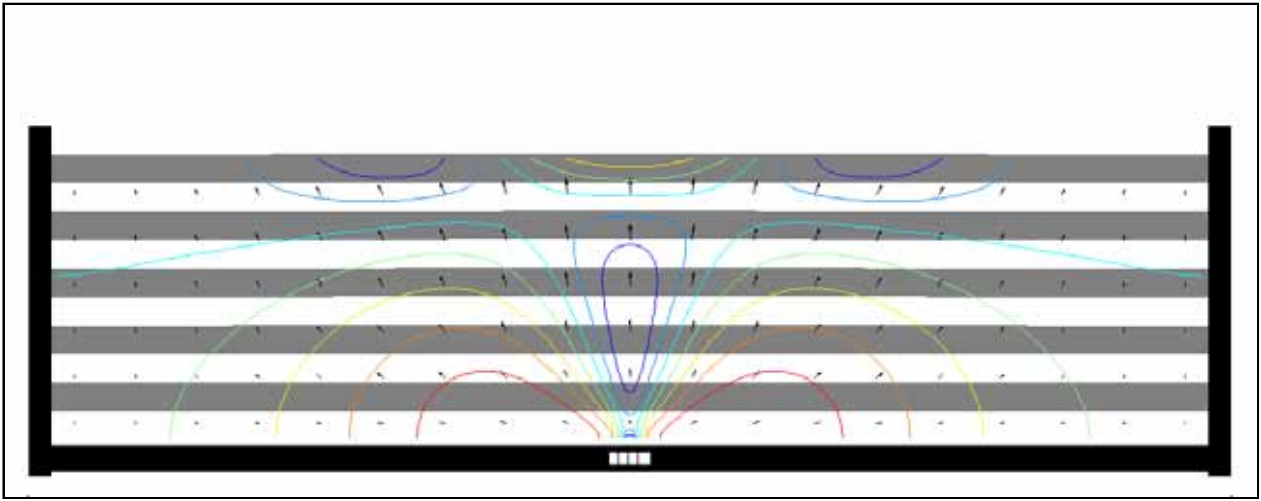

**Pattern II:**

$\dfrac{\sigma_H}{\sigma_V} = 3$, $\varphi = 33^o$, $\upsilon = 0^o$, $\dfrac{C}{\sigma_V} = 0.1$, $\dfrac{\sigma_T}{\sigma_V} = 0.1$, $\nu = 0.3$, $\dfrac{G}{\sigma_V} = 10^7$, $\dfrac{L}{h} = 4$, $\dfrac{w}{h} = 10^{-1}$, $\alpha = 1$. The first frame in the animation corresponds to the elastic solution at failure onset $p = p^c$, the last frame corresponds to the case when $p = 1.8 p^c$. The pore pressure increases linearly with time during the animation.

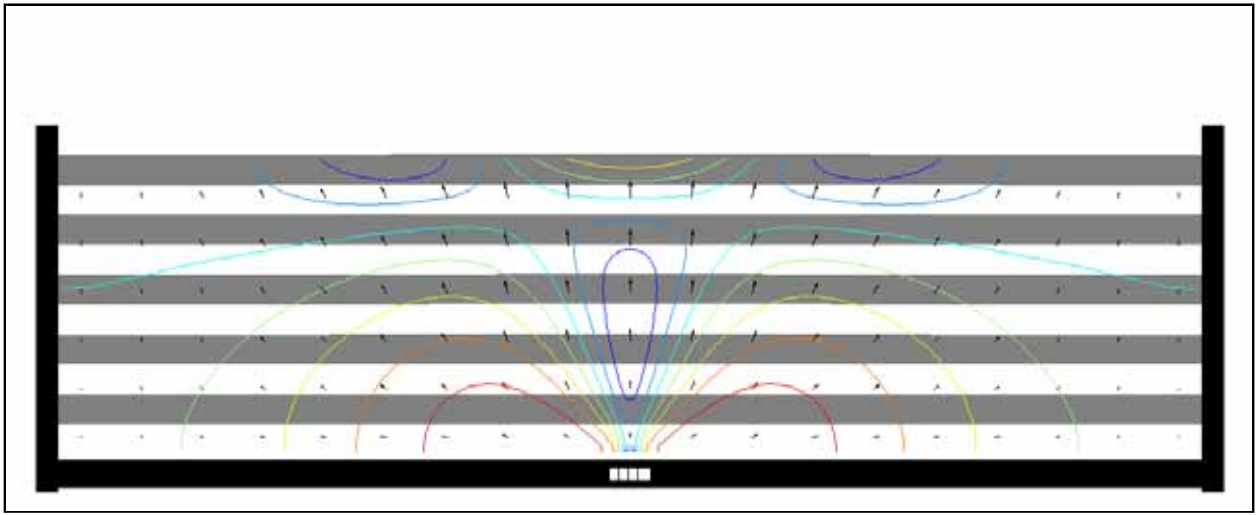

**Pattern III:**

$\dfrac{\sigma_H}{\sigma_V} = 0.1$, $\varphi = 33^o$, $\upsilon = 0^o$, $\dfrac{C}{\sigma_V} = 0.8$, $\dfrac{\sigma_T}{\sigma_V} = 0.266$, $\nu = 0.3$, $\dfrac{G}{\sigma_V} = 10^7$, $\dfrac{L}{h} = 4$, $\dfrac{w}{h} = 10^{-1}$, $\alpha = 1$. The first frame in the animation corresponds to the elastic solution at failure onset $p = p^c$, the last frame corresponds to the case when $p = 4 p^c$. The pore pressure increases linearly with time during the animation.

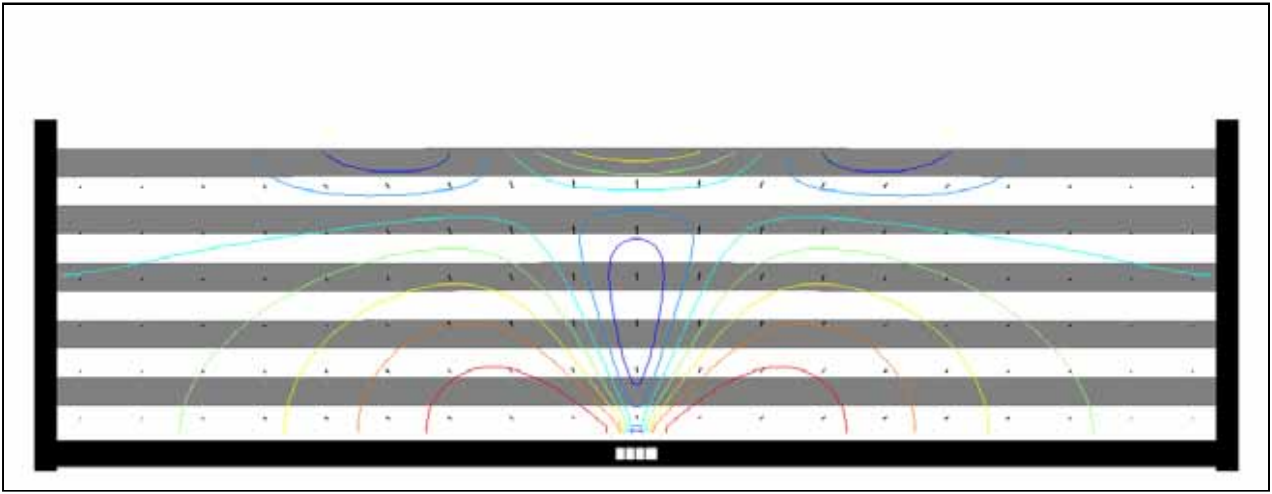

**Pattern IV:**

$\dfrac{\sigma_H}{\sigma_V} = 5$, $\varphi = 33^o$, $\upsilon = 0^o$, $\dfrac{C}{\sigma_V} = 4$, $\dfrac{\sigma_T}{\sigma_V} = 0.266$, $v = 0.3$, $\dfrac{G}{\sigma_V} = 10^7$, $\dfrac{L}{h} = 4$, $\dfrac{w}{h} = 10^{-1}$, $\alpha = 1$. The first frame in the animation corresponds to the elastic solution at failure onset $p = p^c$, the last frame corresponds to the case when $p = 3.2 p^c$. The pore pressure increases linearly with time during the animation.

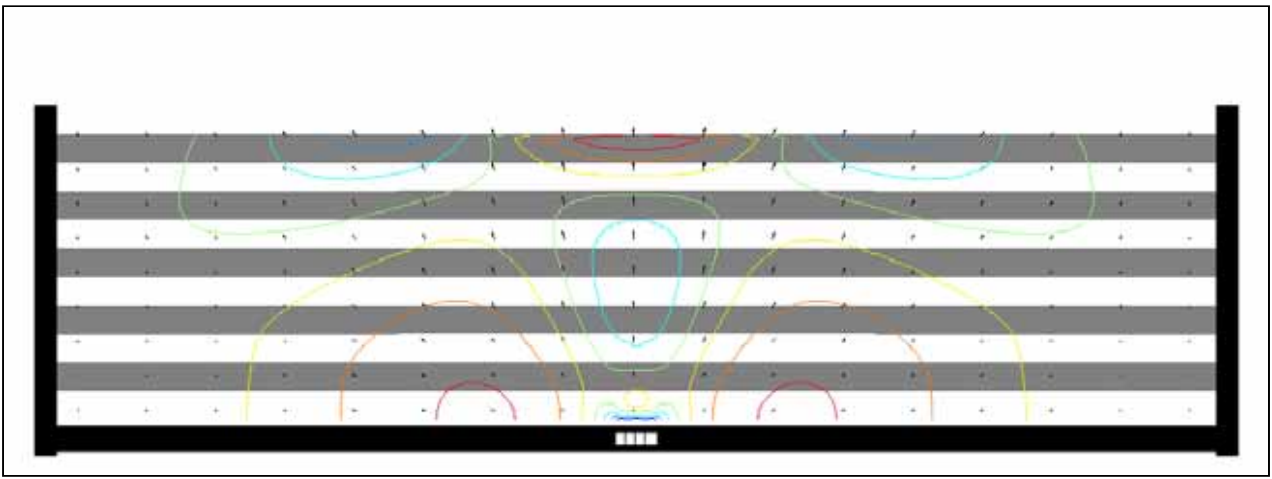

**Pattern V:**

$\dfrac{\sigma_H}{\sigma_V} = 0.1$, $\varphi = 33^o$, $\upsilon = 0^o$, $\dfrac{C}{\sigma_V} = 0.8$, $\dfrac{\sigma_T}{\sigma_V} = 10^{-8}$, $v = 0.3$, $\dfrac{G}{\sigma_V} = 10^7$, $\dfrac{L}{h} = 4$, $\dfrac{w}{h} = 10^{-1}$, $\alpha = 1$. The first frame in the animation corresponds to the elastic solution at failure onset $p = p^c$, the last frame corresponds to the case when $p = 16 p^c$. The pore pressure increases linearly with time during the animation.

**References**

Biot, M. A. (1941), General theory of three-dimensional consolidation, *Journal of Applied Physics, 12,* 155-164.

Goodier, J r. J.N., and Hodge, P.G., (1958), *Elasticity and Plasticity: The Mathematical Theory of Elasticity and The Mathematical Theory of Plasticity,* 152 pp., John Wiley & Sons, New York.

Lavrent'ev, M. A., and Shabat, B. V. (1972), *Methods of Theory of Complex Variable Functions*, 736 pp., Nauka, Moscow.

Kolosov, G. V., (1909), *On the Application of the Theory of Functions of a Complex Variable to a Plane problem in the Mathematical Theory of Elasticity*, Ph.D. diss., Dorpat University (in Russian).

Lebedev, N. N., (1937), Thermal Stresses in the Theory of Elasticity, ONTI, Moscow-Leningrad, 55-56 (in Russian).

Muskhelishvili, N.I., (1977), *Some Basic Problems in the Mathematical Theory of Elasticity*, 768 pp., Springer, Berlin.

Paterson, M. S., and Wong, T.-F. (2005), *Experimental Rock Deformation-The Brittle Field*, 346 pp., Springer, Berlin.

Rice, J. R., and M. P. Cleary (1976), Some basic stress-diffusion solutions for fluid-saturated elastic porous media with compressible constituents, *Reviews of Geophysics and Space Physics, 14*, 227-241.

Timoshenko, S. P., and Goodier, J.N. (1982), *Theory of Elasticity*, 608 pp., McGraw-Hill, New York.